# Competing advection decelerates droplet evaporation on heated surfaces

**Abhishek Kaushal [a], Vivek Jaiswal [a], Vishwajeet Mehandia [a]** and **Purbarun Dhar [b, *]**

[a] Department of Mechanical Engineering, Indian Institute of Technology Ropar, Rupnagar–140001, India

[b] Department of Mechanical Engineering, Indian Institute of Technology Kharagpur, Kharagpur–721302, India

*Corresponding author:*

**E–mail:** purbarun@mech.iitkgp.ac.in ; purbarun.iit@gmail.com

**Tel:** +91-3222-28-2938

## Abstract

In this article we report the atypical and anomalous evaporation kinetics of saline sessile droplets on surfaces with elevated temperatures. In a previous we showed that saline sessile droplets evaporate faster compared to water droplets when the substrates are not heated. In the present study we discover that in the case of heated surfaces, the saline droplets evaporate slower than the water counterpart, thereby posing a counter-intuitive phenomenon. The reduction in the evaporation rates is directly dependent on the salt concentration and the surface wettability. Natural convection around the droplet and thermal modulation of surface tension is found to be inadequate to explain the mechanisms. Flow visualisations using particle image velocimetry (PIV) reveals that the morphed advection within the saline droplets is a probable reason behind the arrested evaporation. Infrared thermography is employed to map the thermal state of the droplets. A thermo-solutal Marangoni based scaling analysis is put forward. It is observed that the Marangoni and internal advection borne of thermal and solutal gradients are competitive, thereby leading to the overall decay of internal circulation velocity, which reduces the evaporation rates. The theoretically obtained advection velocities conform to the experimental results. This study sheds rich insight on a novel yet anomalous species transport behaviour in saline droplets.

# 1. Introduction

Thermo-hydrodynamics and species transport in microscale droplets and sprays has been an area of intense interest to the researcher community for the past two decades. Droplet evaporation, an apparently simple phenomenon, yet inherently rich in physics, is encountered in daily life in a wide range of applications. Typical processes and systems include the automobile industry and combustion engines [1], in inkjet printing [2], spray cooling in the metallurgy industry [3], surface coating and texturing [4], manipulation of biological fluids [5], in sol-gel-technology [6], power generation engineering [7-8], etc. Several researchers have been working in the area of evaporation dynamics of droplets owing to its bio-medical applications as well. These applications include nebulizers and inhalers [9], patterning and detection of diseases from blood [10], spray and fumigants, etc. Combustion dynamics of fuels is also an area directly associated with droplet evaporation kinetics [11) and is studied widely.

The physics behind sessile droplet evaporation is an intriguing problem, as it is strongly dependent on the interactions between the droplet fluid molecules, the surrounding gas phase, and the surface involved [12-14]. . Maxwell [15] introduced the mass diffusion mechanism based simple droplet evaporation model, wherein dimensionless numbers were applied to evolve the governing equation and the value of droplet evaporation rate was determined. Picknet and Bexon [16] reported the existence of two distinct modes of evaporation, the constant contact radius (CCR) and the constant contact angle (CCA) modes. Further, strong dependence of evaporation rates on the wetting state was reported. Later, four intermittent modes of evaporation were reported in case of partially wetted substrates [17]. Bourges-Monnier and Shanahan [18] explored the influence of contact angle on the evaporation process and also discussed the four stages of evaporation depending on the roughness of different hydrophilic substrates, for water and n-decane droplets.

Many prior studies on sessile droplet evaporation focussed on the CCR mode of evaporation [16, 19-21]. Popov [22] reported a vapour diffusion model based closed-form expression for the evaporation rates for the entire range of contact angles. Numerous studies to understand the role of surface wettability on the evaporation dynamics have been put forward [23, 24]. Sessile droplet evaporation on superhydrophobic surfaces (SHS) has been observed to follow three major modes of evaporation, viz. the CCA, the CCR, and the mixed mode [25, 26]. Dash and Garimella [27] experimentally investigated the evaporation rate of sessile droplets on SHS with sliding contact line. Many studies have reported the importance of internal hydrodynamics on the evaporation kinetics of sessile droplets. The temperature gradients along the liquid-air interface of the droplet exists due to non-uniform evaporative cooling, leading to surface tension gradients [28, 29], which give rise to thermal Marangoni advection inside the droplet [30-32]. These flows can either counter or aid the capillary flows inside the droplet.

Xu and Lo [33] showed that the Marangoni flows also exist within evaporating pure water droplets; however the advection is very weak. The role of wetting states on the Marangoni flow inside the droplet has also been studied [34, 35]. Apart from the temperature gradients, presence of solute or other solvent phase in the droplet can induce surface tension gradients which results in solutal Marangoni convection. Bennacer and Sefiane [36] reported the presence of multiple vortices at the liquid-air interface during initial period of evaporation, caused by the Marangoni effect (concentration gradient) in water-ethanol droplets. Kang et al. [37] experimentally noted the existence of two stable counter rotating vortices inside a sessile droplet of water-ethanol mixture. The dominance of Marangoni advection over buoyancy driven Rayleigh convection inside aqueous NaCl droplets on hydrophilic [38] and hydrophobic surfaces [39] have been reported in literature. Karapetsas et al. [40] reported the influential role of dispersed particles on the flow dynamics and evaporation rates of the droplets.

Articles discussing the role of thermal properties of the substrate on sessile droplet evaporation dynamics have also been reported. The role of the thermal conductivity of the surface [41, 42] towards altering the internal flow direction and evaporation rate of droplets has been reported. Substrate heating is also known to modulate the evaporation dynamics of sessile droplets, as reported by Girad et al. [43], Saada et al. [44], Carle et al. [45] and Pradhan et al. [46]. Tam et al. [35] reported the presence of two counter rotating vortices inside a droplet placed on a heated SHS. Sobac and Brutin [47] studied ethanol droplet evaporation on heated substrates and proposed a quasi-steady diffusion driven model for the evaporation rates. Kim et al. [48] investigated the influence of the substrate temperature on the flow directions of the solute particles inside a sessile droplet during evaporation process. Recently, the present authors (Kaushal et al. [49]) showed experimentally and theoretically that saline sessile droplets evaporate faster compared to water droplets when placed on non-heated surfaces (hydrophilic and SHS). It was shown that the interfacial shear generated by the internal solutal hydrodynamics aided the external Stefan flow, leading to augmented evaporation.

Taking cue from the previous study [49], this study extends to understand the evaporation kinetics of sessile saline droplets with substrate heating. Experimental and analytical approach to understand the evaporation behaviour with induced thermal gradient within the droplets for different wettability substrates (on hydrophilic and SHS) has been presented. Anomalous evaporation characteristics are noted when compared with the previous study. While the saline sessile droplets evaporated faster compared to the water droplets for non-adiabatic cases, the converse is noted for the heated surfaces. It is observed that the saline droplets evaporate at slower rates than its water counterpart at the same surface temperature. Particle Image Velocimetry (PIV) studies were carried out to investigate the internal advection dynamics and the alteration in the advection behaviour within saline droplets on heated substrates. To understand the dynamics of the thermal Marangoni and Rayleigh convection and the solutal Marangoni advection inside the droplet and their influence on the stability of the internal circulation behaviour, a scaling analysis based mathematical formulation is presented. From the analysis it is obtained that individually, both the thermal and solutal Marangoni advection in augmented by the heated surfaces. However, the two advection components compete and oppose one another, leading to reduction in the overall advection strength. This in turn leads to reduced evaporation rates. The theoretical velocity values obtained are found to be in good agreement with the PIV observations.

## 2. Materials and methods

In the present studies, Sodium Iodide (NaI) salt (procured from Merck, India, and used as obtained) solution in de-ionised (DI) water is employed. The salt is selected based on the results by the authors from literature [49]. Experiments were performed on substrates of two different wettabilities. For hydrophilic surface, sterile glass slides were used, which were cleaned with acetone and dried overnight. The SHS are synthesized by spray coating (Rust Oleum industrial spray, USA) the sterile glass slides and drying overnight in oven. The surface properties of the substrates used are tabulated [49] in the table 1. To measure the ambient temperature and humidity, a digital thermometer and a digital hygrometer with a sensing probe is used. The sensing probe is placed ~ 20 mm away from the evaporating droplet. For all the experiments, the ambient conditions were noted as 25 ± 2 ºC temperature, and relative humidity of 50 ± 5%. The initial temperature of the droplet is ensured as 25 ºC before placing on the substrate.

The substrates were placed on a metallic plate heater made up copper, which is connected to a T-type thermocouple and a digital heating unit controller (refer fig. 1) with thermostat controller to control the plate temperature. The heating unit is calibrated with respect to the temperature drop across the glass slides and in the present experiments, the temperature of the substrate is varied to three different values with respected to the ambient temperature (25 ºC), viz. Δ0, Δ10 and Δ30, where thermal difference Δ0 is the ambient case. The difference above 30 is not studied as the evaporation rate is augmented to a very large extent at higher temperatures, which leads to rapid crystallization induced anomalies in the saline droplets, as well as the reduced droplet life-time prevents error free velocimetry. No significant change (not greater than ~10%) in the relevant physical properties of the fluid (except for the viscosity and surface tension) is noted due to rise in the substrate temperature. The reproducibility of the observations is verified by repeating each experiment thrice, and the average of the observations is considered.

An experimental setup (shown in fig. 1) assembled in-situ is utilized in the present study. The droplets were dispensed carefully using a digitized droplet dispensing mechanism (Holmarc Opto-mechatronics, India), capable of dispensing droplets accurate to 0.1 μl volume. The sessile droplets are dispensed carefully so as to ensure that the droplets do not undergo any spreading or retraction due to the pumping effect or capillarity within the needle. The volume of the droplets used is 20 ± 0.5 μL, chosen such that the contact diameter of the sessile droplet is less than the capillary length scale for water (the associated Bond number is maintained < 1, such that gravity effects are not important). A monochromatic, CCD (charged couple diode) camera (Holmarc Opto-mechatronics, India), attached to a long-distance microscopic lens, and capable of recording at 30 fps at 1 megapixel resolution, is used to record the evaporation process. The camera is mounted on a 3 axis translational stage for focusing. A brightness-controlled LED array light source (DPLED, China) was used as an illumination source for the camera.

**Table 1**: The liquid-gas and solid-gas components of surface energy and static contact angles for the substrates used in the experiments (for glass, $\sigma_{sg} = 0.375$ J/m$^2$):

| Substrate | $\sigma_{sl}$ (J/m$^2$) | $\sigma_{lg}$ (J/m$^2$) | Static contact angle |
|---|---|---|---|
| Glass | 0.316 | 0.0728 | $40^\circ \pm 3^\circ$ |
| SHS | 0.441 | 0.0728 | $155^\circ \pm 3^\circ$ |

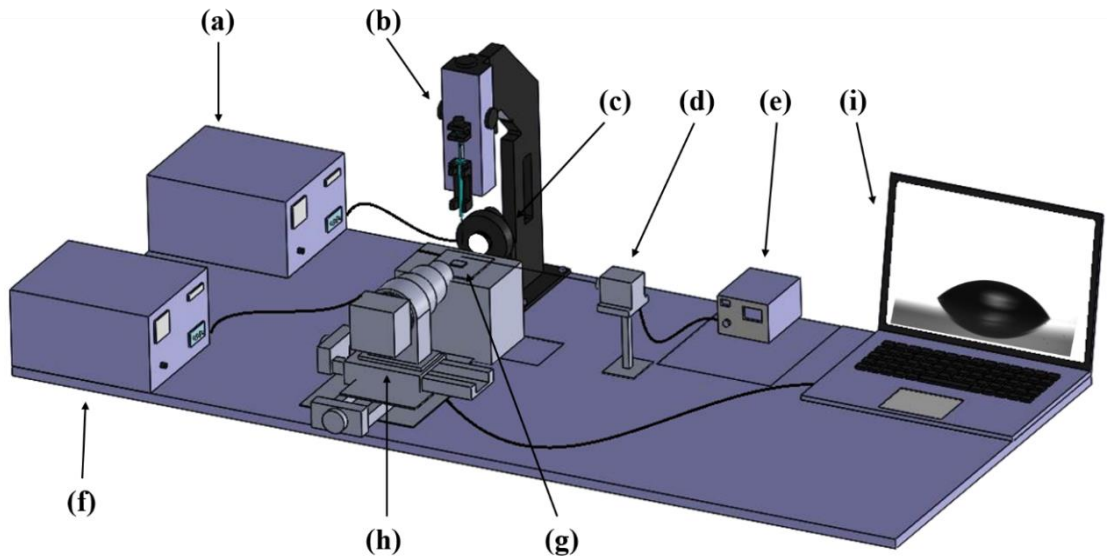

**Figure 1:** The model of the experimental setup, consisting of (a) droplet dispenser controller unit, (b) droplet dispensing mechanism with Hamilton syringe, (c) LED backlight assembly, (d) laser with sheet optics assembly (not shown), (e) laser controller (f) controller connected to the copper plate heating unit, (g) substrate with droplet on heating mechanism, (h) CCD camera with long distance microscope lens and three-axis positioning system, (i) computer for data acquisition and camera control. The components (b, c, d, g and h) are enclosed in an acrylic chamber and mounted on a vibration-free table.

The whole experimental setup was kept on a vibration free table and housed within an acrylic chamber to eliminate external disturbances. An open source software (ImageJ) was used for image processing by using macro subroutines, where each image was first converted into binary and the geometric parameters of the droplet were extracted. The visualization of the internal flow behaviour and quantification of the advection velocities is done using particle image velocimetry (PIV). Neutrally buoyant, fluorescent particles (~10µm diameter of polystyrene, Cospheric LLC, USA) were used as the seeding particles. A continuous wave laser (532 nm wavelength, 10 mW peak power) was used as the illumination source (Roithner GmbH, Germany) for the PIV. The PIV was done at 30 fps and ~120 pixels/mm resolution. A cylindrical lens is used to generate a laser sheet of thickness ~0.5 mm to focus the vertical mid-plane of the droplet. This is done in case of the droplets on the SHS as PIV of the vertical plane is of good resolution.

For the case of hydrophilic substrate, the droplet height is very low, and hence PIV of the vertical plane of the droplet is of poor resolution. Consequently, top view micro-PIV studies are done using a fluorescent microscope. A CMOS monochrome camera (Sony Corpn.) is used at 30 fps and 10x optical zoom of the microscope. The microscope objective is focused at the horizontal mid-plane of the droplet. For the micro-PIV, graphite nanoflakes (0.3–0.5 μm size) have been used as the seeding particles. Using image processing, the images were later binary inverted such that the particles were converted to white pixels and the fluid domain was converted to black. The open-source PIV code PIV-Lab was used for the velocimetry analysis. A four pass, cross-correlation algorithm has been employed in the post processing, with consecutive interrogation window sizes of 64, 32, 16, and 8 pixels, to obtain high signal to noise ratio. Standard noise reduction and contrast enhancing pre-processing algorithms are employed to enhance peak locking. Infrared thermography (at 4X thermal zoom) has been used to determine the thermal gradients and temperature distribution within the evaporating droplets (FLIR T650sc infrared camera). It employs an infrared detector of resolution 640 x 512 pixels and has an accuracy of ±0.3 $^{o}$C in the temperature range of 0-100 $^{o}$C.

## 3. Results and Discussions

### 3. A. Role of substrate temperature on evaporation kinetics

The evaporation rate in the sessile droplets is likely to enhance with temperature [43]. Figure 2 presents the time snap array of the evaporating sessile droplets of 0.1M concentration at different substrate temperatures. It has been found in the literature as well as observed experimentally that the droplets on the hydrophilic substrate (Fig. 2(a)) evaporate faster as compared to that on superhydrophobic substrate (Fig. 2(b)). The influence of the different thermal gradients on both the substrates produce an observable changes in the evaporation kinetics of droplets. In Fig. 2(a), it is observed that the droplets on hydrophilic substrate remain pinned during most of the evaporation process and only height and contact angle of the droplet decreases. While on the other hand, in SHS (fig. 2(b)), the contact angle decreases at comparatively lower rate as compared to contact diameter in the initial stage of evaporation. The volume of the droplet is observed to decrease significantly and the decrease is direct function of the increase in temperature of the substrate. However the presence of salt in the droplet solution is observed to behave differently than the actual behaviour. From the previous study [49] it is reported that the presence of salt is likely to increase the evaporation kinetics in the sessile droplets on hydrophilic surface as well as on SHS (lower concentrations). But in the present scenario, the presence of salt in the solution decreases the evaporation rate of the droplets in case of elevated substrate temperatures on both hydrophilic substrate and SHS and this decrease is the direct function of the concentration of the salt in the solution.

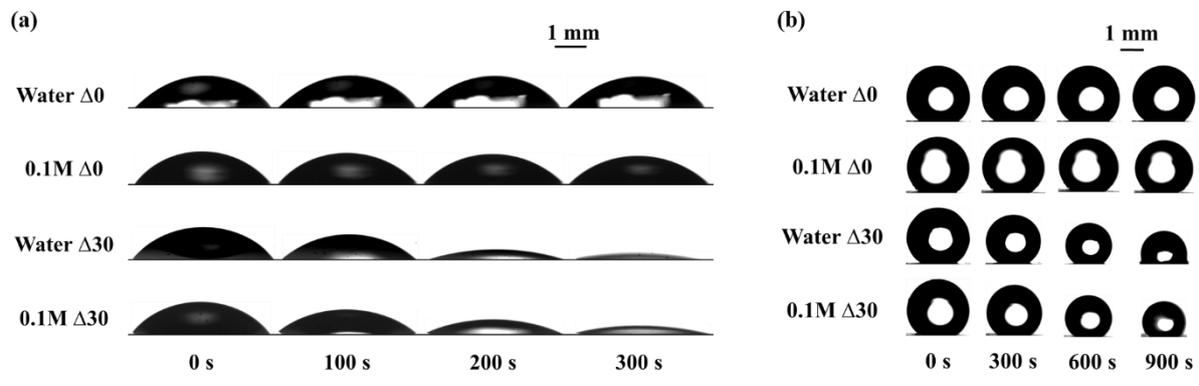

**Figure 2:** Time–snap array of the evaporating sessile droplets of water droplets and 0.1 M NaI solution on (a) hydrophilic substrate, and (b) SHS, under different thermal gradients (Δ0 and Δ30).

Figure 3 illustrates the transient variation of the non-dimensional characteristic volume V* ($V^* = \left(\dfrac{V}{V_0}\right)^{2/3}$, where V is the instantaneous volume and $V_0$ is the initial volume of the droplet) for different thermal conditions. In case of hydrophilic substrate (fig. 3(a)), for the Δ0 case, the evaporation rate of the droplets is observed to increase with increase in the concentration of the salt. The saline droplets thus evaporate faster than the water counterpart. However, strangely, an anomaly is noted when the surface temperature is higher than the ambient. With rise in the substrate temperature, for a particular salt concentration, the presence of salt is observed to reduce the droplet evaporation rate compared to its water counterpart. And this effect is more dominant with increase in the temperature of the surface compared to the ambient. Similarly in the case of SHS, at Δ0, the droplet evaporation rate first increases with salt concentration but then decreases, which has been explained by the present authors earlier [49]. However, distinct differences between the different cases at the ambient conditions are noticeable. As the substrate temperature is enhanced to Δ10 condition, the evaporation curves all collapse towards one another, and the apparent differences between water and saline droplets disappear. This is also in stark contrast to the ambient observation. Thereby the observations reveal that the conjugate effect of salt and substrate heating lead to atypical and counter-intuitive

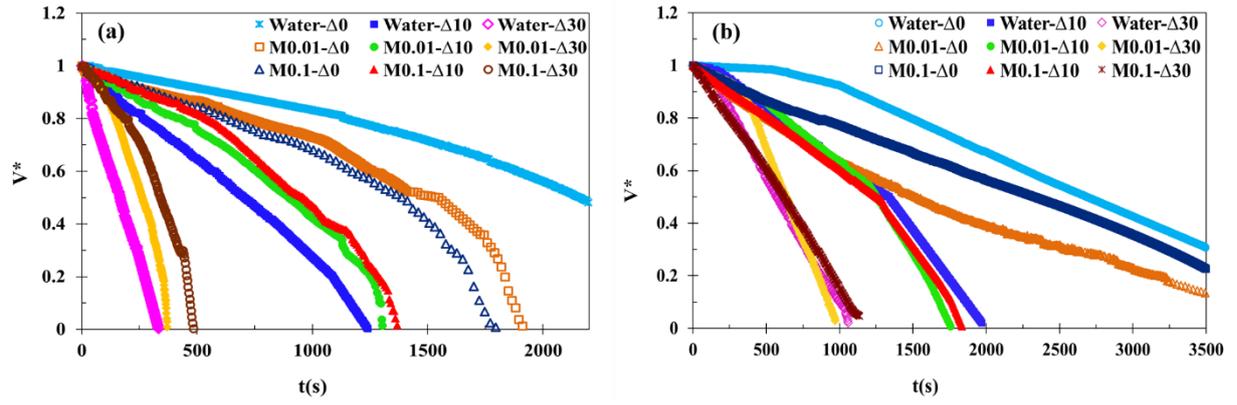

**Figure 3:** Behaviour of V* with time during the evaporation of water and NaI solution droplets for different temperature conditions on (a) hydrophilic substrate and (b) SHS.

Interfacial tension plays an important role in the evaporation dynamics of sessile droplets. The static contact angle of the droplet depends on the equilibrium of the surface tension forces, which in turn dictates the evaporation modes. Figure 4 illustrates the temporal variation of the normalized contact angle (with respect to the initial static contact angle) of the droplets. It is observed that at ambient conditions, the contact angle on hydrophilic substrate (fig. 4 (a)) decreases almost linearly from the very initial stage (CCR mode), followed by the mixed mode, and then remains constant (CCA mode) towards the end of the evaporation process. However, at Δ10 and Δ30, the behaviour is modulated. The mixed mode of evaporation is observed in the later stage, with very little or no CCA mode at all. While the presence of the salt tends to arrest the evaporation kinetics at elevated temperatures, the similar anomalous effect on contact angle variation is also visible. On the contrary, on SHS and ambient case, the contact angle reduces slightly in the initial stages (CCA mode), followed by mixed mode where the droplet retracts itself over and over [49], and then reduces rapidly towards the end (CCR mode). With rise in the substrate temperature, it is observed that the droplet tends to remain in the mixed regime for most of its evaporation time, followed by a sudden jump to the CCR regime. Here also the substrate temperature causes the curves to collapse and the differences in the ambient case are no longer prominent.

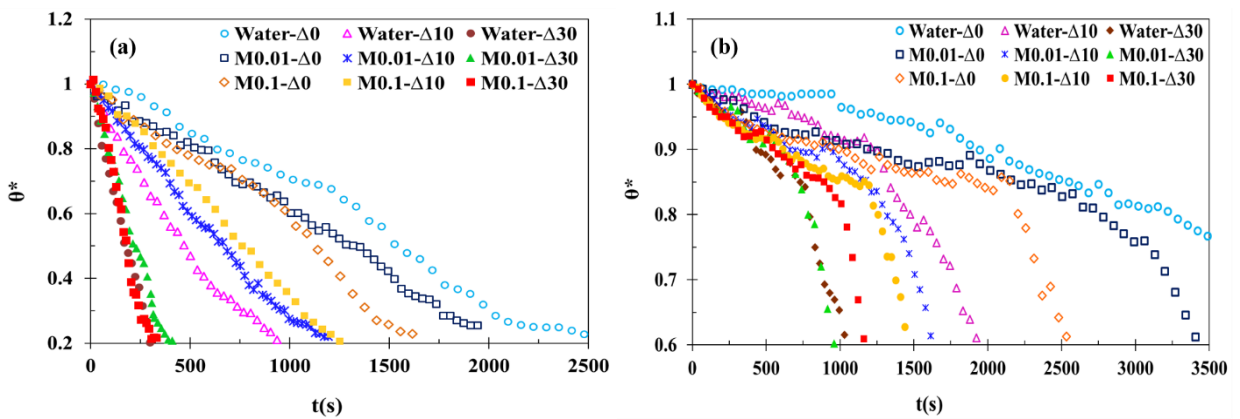

**Figure 4:** Variation of the non-dimensional contact angle θ* with time for water and NaI solution droplets on (a) hydrophilic surface (b) SHS for different substrate temperatures.

The transient response of the non-dimensional contact diameter (with respect to initial contact diameter) of the droplet has been shown in figure 5. On hydrophilic substrate (fig. 5(a)), it is noted that in ambient case, the CCR regime is reduced with addition of salt, when compared to water droplet. However, with increase in the substrate temperature, it is observed that the CCR regime is extended with salt compared to the water droplet. This behaviour also provides another instance of anomaly with respect to the ambient case observations, when the saline droplet evaporates at elevated substrate temperatures. On the SHS (fig 5(b)), it is seen that for the ambient case, the behaviour first deviates largely from the water case, but tends towards the water case at higher salt concentration. This behaviour has been noted and explained by the authors in the previous literature [49]. However, for the Δ10 and Δ30 case, it is seen that this unique behaviour is completely absent, and the saline droplets show consistent change in the contact angle with respect to concentration. The anomaly is furthered by the fact that this behaviour is exactly opposite to the observations in fig. 5 (a).

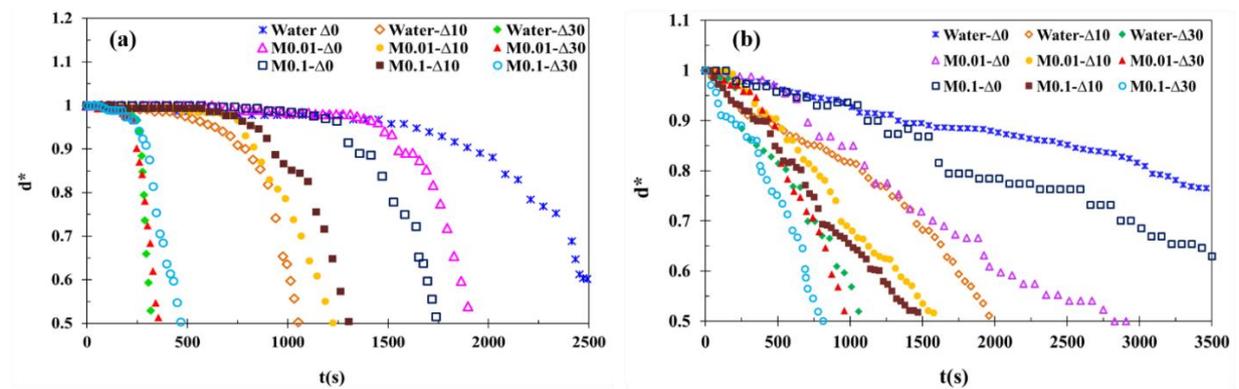

**Figure 5:** Variation of the non-dimensional contact diameter d* with time for water and NaI solution droplets for different substrate temperatures on (a) hydrophilic substrate and (b) SHS.

Having reported the observations, it is now essential to probe and discuss the mechanisms which are responsible towards such anomalous behaviour. Changes in surface tension and role of diffusion mediated evaporation have been shown to bring about consistent changes in the evaporative rates [46], and thus can be safely excluded from the discussion on mechanisms behind such atypical responses. It has been noted from the experimental image processing that the contact line recedes differently on both the substrates during the whole evaporation process, and the modulation of the same with thermal stimulus is also different. It is also interesting to find the effect of substrate temperature on the contact line receding velocity of the droplet (fig. 6). It is observed that the contact line does not start receding in the initial stage of evaporation even at higher temperatures both in water as well as salt based droplets. But with rise in the substrate temperature, the contact line receding velocity increases in both the cases and also is direct function of the substrate temperature. It has also been observed that with addition of salt, contact line receding velocity increases in magnitude as compared to the water droplet at ambient temperature (shown in the fig. 6(a)) but decreases on substrate with higher temperatures.

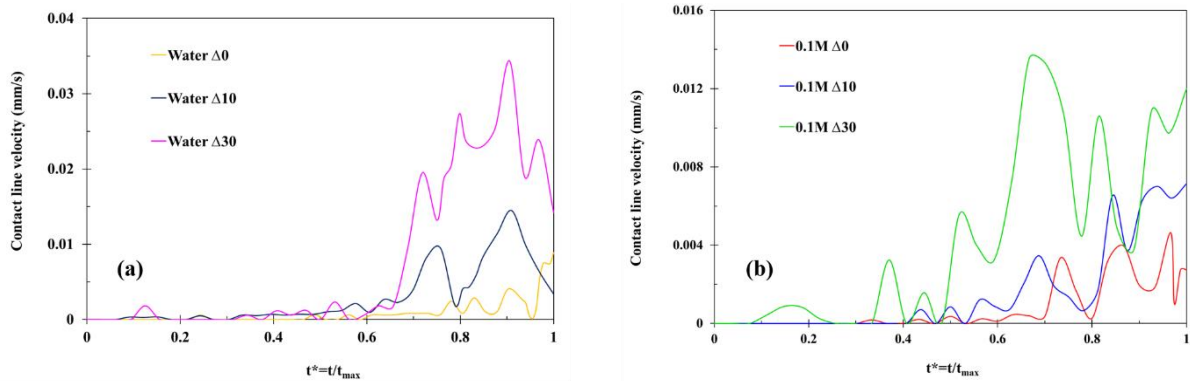

**Figure 6:** Variation in the contact line receding velocity with non-dimensional time $t^*$ (non-dimensionalized by the droplet life-time) on hydrophilic substrate at different substrate temperatures in (a) water droplet and (b) 0.1 M NaI saline droplet

### 3. B. Behaviour of the internal hydrodynamics

The presence of solvated components and inclusions is known to modify the evaporation kinetics of droplets by augmenting the internal hydrodynamics of the droplets [49]. This internal advection in turn shears the liquid-gas interface, which entrains the external vapour diffusion layer with the surrounding ambient phase. This aids the Stefan flow surrounding the droplet, which leads to enhanced species transport. Thereby, it is necessitated that the internal hydrodynamics be probed to shed possible light on the observed anomalies. To this end, particle image velocimetry (PIV) was done for all cases within the first few minutes of the evaporation process. The PIV studies were performed in the initial regimes to prevent the effect of change in flow dynamics due to change in saline concentration within the droplet (the bulk concentration changes by > 10% even in that short time frame for elevated temperatures).As discussed in the methods section, the imaging was done in the vertical mid-plane for the SHS droplet, and at the horizontal mid-plane for the hydrophilic droplet.

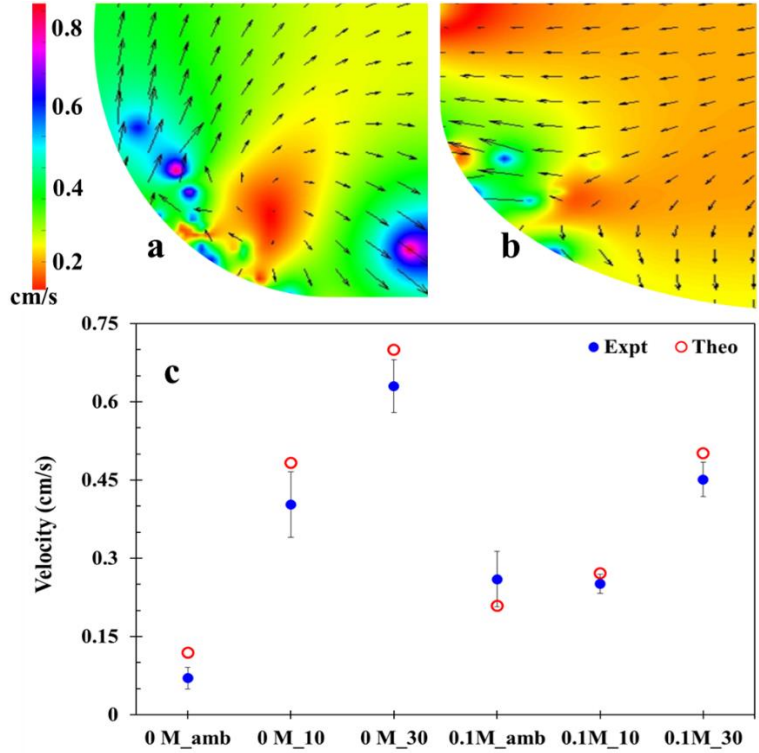

**Figure 7:** Time averaged velocity contours and vector fields for (a) water droplet and (b) 0.1 M NaI droplet, on hydrophilic surface, for at Δ10 case. A quadrant of the droplet has been considered for the velocimetry. (c) Comparison of the experimentally obtained spatio-temporally averaged advection velocities and the predicted theoretical velocities for the hydrophilic surface..

Figure 7(a) illustrates the temporally averaged velocity contours and vector fields for water droplet and figure 7(b) illustrates the same 0.1 M saline droplet, both on hydrophilic surface of Δ10 case. The temporally averaged velocity vectors and contours are obtained from analysis of 1000 consecutive image frames. With increase in the substrate temperature, the strength of advection within the water droplet has enhanced significantly compared to the ambient case (not illustrated), which complies with previous report [46]. However, in case of the saline droplet on heated substrate, the strength of the internal advection is reduced compared to the corresponding water case (fig. 7 (a)). Furthermore, it is interesting to note that the while the direction of internal circulation is dissimilar in case of the water and saline droplets on heated surfaces. At the horizontal mid-plane, the water droplet exhibits inward internal advection, whereas the saline droplet exhibits outward internal advection currents.

On the SHS (fig. 8), increase in substrate temperature has been shown [46] to enhance the internal advection in water droplets and similar results were obtained for water in the present case (fig. 8 (a)). Also, the strength of internal circulations in the droplet on the SHS is higher in magnitude compared to the droplet on hydrophilic substrate, due to the hydrophobicity induced Marangoni stress based currents in the SHS droplet [49]. It has been observed in the case of saline droplet that at higher substrate temperatures, the advection currents within the droplet are largely diminished in magnitude compared to the water droplet under same thermal conditions. However, the opposite is true in case of the ambient evaporation case (fig. 8 (c)). Thereby, it is noted from the PIV exercises that increase in surface temperature leads to increase in internal advection strength in case of water droplets. In the ambient condition, saline droplets show higher internal advection velocity than water droplets. However, in the event of heated surface, the saline droplets show diminished internal circulation velocity compared to the water droplet at same temperature. The PIV findings thus support that an anomalous and counter-intuitive phenomenon is discovered in the present study. Intuition and water droplet observations suggest enhanced internal advection in case of heated surfaces; however, the opposite is noted in saline droplets. Thus, a deep probing of the reason behind the same is necessitated.

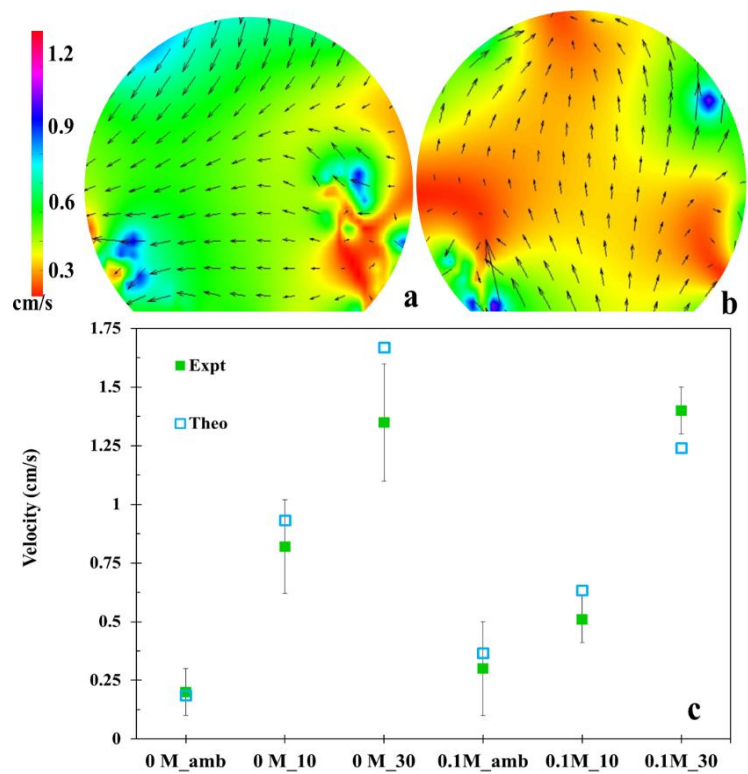

**Figure 8:** Time averaged velocity contours and vector fields for (a) water droplet and (b) 0.1 M NaI droplet, on SHS, for Δ10 condition. (c) Comparison of the experimental spatio-temporal averaged velocities with the theoretically obtained velocities.

### 3. C. Dynamics of the internal thermal advection

In the previous report by the authors [49], the internal advection is found to be major factor behind the modulated evaporation dynamics of the saline droplet, with the solutal Marangoni effect as the major cause behind the advection. With rise in the substrate temperature, the thermal advection is likely to enhance. The temperature inside the droplet keeps on changing during the evaporation process, with centre of the droplet to be at minimum temperature due to evaporative cooling and the vapour-liquid interface at the maximum. Thermographic pictures of the droplets of 0.1M concentration taken at initial time period of evaporation process at different substrate temperatures are shown in Figure 9(a). The difference in the thermal profile due to presence of the thermal gradient in substrtae temperatures can be clearly observed. Also the various thermal regimes can be seen within the droplet due to evaporative cooling. Figure 9(b) and 9(c), illustrate the non-dimensional temperature profile inside the droplet along the non-dimensional radius r* (r*= r/$r_0$, $r_0$ is the initial radius of the droplet) of the droplet. The temperature of the droplet is non-dimensionalized as $T^* = \dfrac{T - T_{min}}{T_{max} - T_{min}}$ where $T_{min}$ is minimum temperature and $T_{max}$ is the maximum temperature inside the droplet. These results are based on the infrared images data taken during the initial five minutes of the evaporation process to overlap with the PIV timeframe. The distribution obtained in the data is due to the evaporative cooling of the droplet's bulk.

Figure 9(b) reports the thermal distribution inside the droplet on the hydrophilic substrate. It can be observed that with increase in the concentration of the droplet, temperature profile inside is deviating from the linear behaviour near the centre of the droplet [49]. The drop in the droplet temperature near the centre is also noticeable at the elevated temperatures. But towards the periphery (droplet-vapour interface), the temperature profile appears to follow linear behaviour in all cases. On SHS (fig 9(c)), there is sudden drop in the temperature profile from the linear behaviour near the centre of the droplet followed by some linear behaviour in all the cases and then again deviates towards the periphery. Also the higher concentrated droplet shows more deviation from linear behaviour.

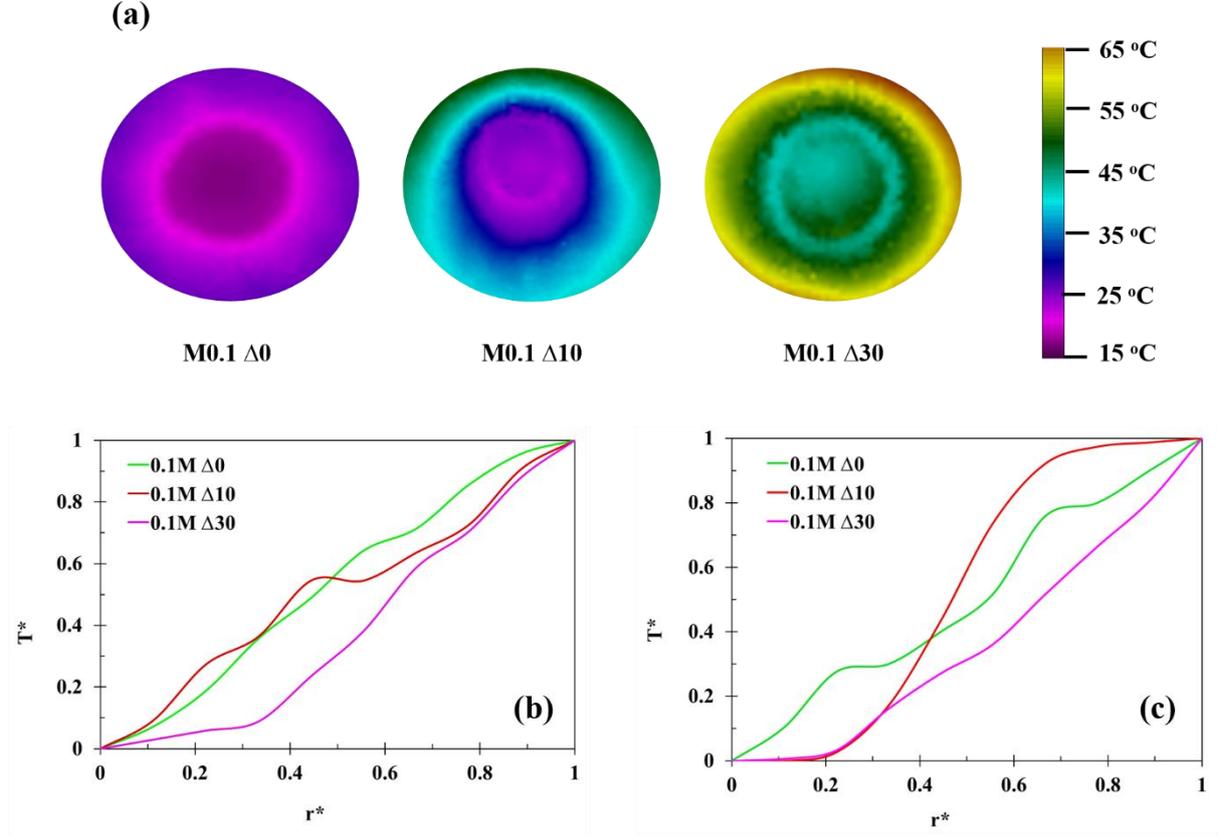

**Figure 9:** (a) Infrared images of the 0.1 M droplets on hydrophilic substrate at different substrate temperatures. Variations in the non-dimensional temperature T* within the droplet along the non-dimensional radius r* at different substrate temperatures on (b) hydrophilic substrate and (c) SHS.

The role to thermal Marangoni advection in case of elevated substrate temperatures is understood using a mathematical scaling model, proposed along the lines of the previous report by the authors [49]. The balance principle is appealed to for the evaporating droplet, and the energy transport components, i.e., due to heat diffusion, due to thermal advection [50, 51], and the energy transport at droplet-substrate interface are modelled. The resulting equation is expressed as

$$\dot{m}h_{fg} = 2kA_c \Delta T_M \frac{\cot(\theta/2)}{d_c} + \rho C_p u_M \Delta T_M A_c \sec^2(\theta/2) - A_c \sigma \dot{\theta} \sin\theta \quad (1)$$

Where, the left hand side represents the energy flux due to evaporation, and the terms of the right hand represent the modes of thermal transport in the droplet [49]. This equation can be further extended to the form [49]

$$\rho h_{fg} d_c^2 \dot{\theta} \sec^4(\theta/2) = 16k\Delta T \cot(\theta/2)\left\{Ma \sec^2(\theta/2) - 8Ma\left(\frac{Ja}{Ca}\right)\sin\left(\frac{\theta}{2}\right)\cos\left(\frac{\theta}{2}\right)\right\} \quad (2)$$

Where, $Ma_T$, $Ca$, and $Ja_e$ represent the associated thermal Marangoni number, the Capillary number and the evaporation Jacob number, respectively [49]. Mathematically, the numbers can be expressed in terms of droplet parameters as

$$Ma_T = \frac{\sigma_T \Delta T_M d_c}{2\mu\alpha} tan(\theta/2) \tag{3}$$

$$Ca = \frac{\mu d_c \dot{\theta}}{4\sigma_T} sec^2(\theta/2) \tag{4}$$

$$Ja_e = \frac{\mu \dot{\theta}}{\rho h_{fg}} \tag{5}$$

Where $d_c$, $\theta$, $\dot{m}$, $h_{fg}$, $k$, $C_p$, $\rho$, $\sigma$, $\mu$ and $\alpha$ denote the contact diameter, contact angle, rate of evaporative mass loss, the enthalpy of vaporization, the thermal conductivity of the liquid, specific heat of the liquid, density of the liquid, fluid surface tension, viscosity of the liquid and thermal diffusivity of the liquid, respectively.

The average internal circulation velocity due to thermal gradient $u_m$ is scaled as $u_M = \frac{\sigma_T \Delta T_m}{\mu}$ [55, 59], where $\sigma_T$ is the gradient of surface tension with temperature [40], and $\Delta T_M$ represents the temperature difference inside the droplet occurred due to evaporative cooling (in ambient case) or due to the imposed substrate heating. In all cases, the magnitude of $\Delta T_M$ is obtained from image processing of the thermography images obtained during the experiments. $\Delta T_M$ causes the change in the surface tension of the liquid and causes internal thermal gradients, which drives the thermal Marangoni currents and advection due to thermo-viscous effect . The values of the Capillary and Jacob numbers describe the influence of the surface characteristics on the evaporation kinetics. Due to imposed thermal gradients in case of elevated substrate temperatures, buoyancy driven Rayleigh advection may play a role in augmenting the internal thermal advection. To obtain the associated Rayleigh number (on the liquid side) is expressible as [55, 59]

$$Ra = \frac{d_c^2}{8\alpha} \left( \frac{\rho g \beta d_c h_{fg} \dot{\theta}}{\mu C_p} sec^2(\theta/2) \right)^{1/2} \tag{6}$$

Where, $\beta$ the thermal expansion coefficient and g is the acceleration due to gravity. The increment in the evaporation rates due to Rayleigh convection in the gaseous phase is weaker, and hence [53].

To understand the dominance of the thermal Marangoni and Rayleigh convection over each other, and to deduce the effect of enhanced substrate temperature on these governing numbers towards the modulating the internal circulations, a stability analysis has been put forward. As proposed by Nield [50] and Davis [51], the critical Marangoni ($Ma_c$) and critical Rayleigh number ($Ra_c$) can be used to express the stability of the advection within the system as

$$\frac{Ma}{Ma_c} + \frac{Ra}{Ra_c} = 1 \tag{7}$$

The $Ma_c$ and $Ra_c$ rely on the associated Lewis number ($Le = q_0 d / k$), and their values ~81 and ~1708 respectively [50, 51, 53]. Based on this analysis, a phase plot of $Ma_T$ vs. Ra has been illustrated (figure 10) for the hydrophilic substrate and SHS at different heated cases to understand the dominance of the mechanism responsible for the internal advection. On hydrophilic substrate (fig. 10(a)) and ambient case, the point lies below the stability regimes proposed by Nield and Davis [50, 51] representing that the circulation inside is unstable [49].

But as the substrate temperature is increased (Δ10 and Δ30 case), the points for water shift towards the top and to the right, and moves above the Nield (N) and Davis (D) lines. This indicates that the advection is conditionally stable and higher the temperature, more stable is the circulation behaviour. Further, the cause of this circulation is predominantly thermal Marangoni effect, as the points cross the stability lines in the upward direction. With increase in the temperature, the Ra increases but is very less compared to the $Ra_c$, which signifies negligible role of buoyant forces in inducing the internal circulation. On the SHS, similar behaviour is observed, with the points traversing to the stable circulation regime at elevated temperatures, with the thermal Marangoni effect being the dominant mechanism. The values of $Ma_T$ at elevated temperatures are higher on the SHS as compared to that on hydrophilic substrate.

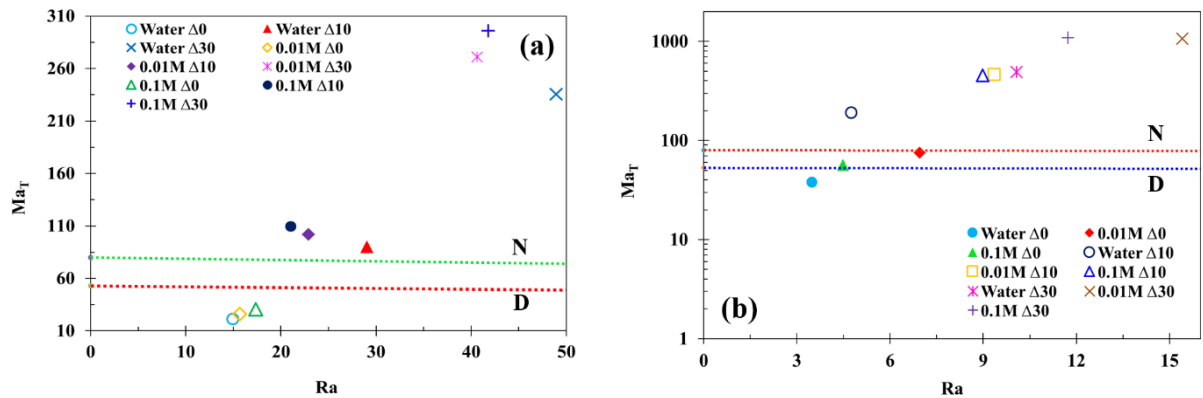

**Figure 10:** Stability plot for $Ma_T$ and Ra for different droplets (a) hydrophilic and (b) SHS at different substrate temperatures. Stability criteria lines proposed by Nield [50] and Davis [51] (labelled as N and D respectively) indicate the regimes for stable internal circulation.

### 3. D. Dynamics of the internal solutal advection

In solvated systems as saline droplets, the solvated ions are preferentially adsorbed and desorbed to the interface and the bulk of the fluid, which leads to modulation of the effective interfacial tension [52-54]. This leads to difference between the values of the solute concentration at the bulk and the interface, which drives the solutal Marangoni advection within such evaporating droplets. Based on the droplet shape analysis, volume measurements, and contact angle mapping during the evaporation process, the transient evolution of the bulk and the interfacial solute concentrations can be determined. The detailed analysis methodology for the same has been described by the authors in previous reports [55-58]. Figure 11 illustrates the comparison of the transient evolution of the interfacial concentration and bulk concentration of the salt during the evaporation, at different substrate temperatures and wettability. Figure 11(a) shows the time evolution of the salt concentration at the bulk and the dynamic interfacial concentration during the evaporation process on hydrophilic substrate, for different heating cases. It is noted that with time, the difference between the bulk and interfacial concentration increases, which complies with reports [49]. Also with increase in the substrate temperature (Δ30 case), the difference in the concentrations reduces as compared to the Δ10 case. This signifies that although the solutal advection is stronger than the thermal counterpart in such saline droplets [49], it is weakened to some extent on heated surfaces. On the SHS (fig. 11(b)), the same trend is observed, and the difference in concentration also reduces with increasing substrate temperature.

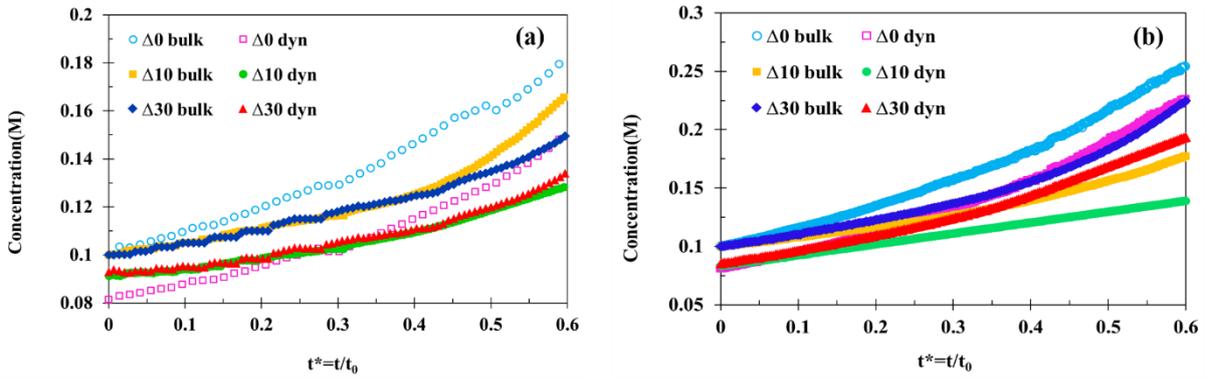

**Figure 11:** The solute concentration at the bulk (denoted as *bulk*), and the dynamic interfacial solute concentration (denoted as *dyn*) during the evaporation of 0.1 M NaI saline droplet at different substrate temperatures, on (a) hydrophilic substrate and (b) SHS. Time on the x-axis is normalized with the total evaporation time.

To map the solutal advection behaviour at elevated substrate temperatures, a similar scaling model based on the species transport mechanisms at play is appealed to [49]. Based on the species transport modes from the droplet and within it, a balance model is expressed as

$$\dot{m} = DA_c \frac{\Delta C}{h(t)} + u_c \Delta C A_s + \frac{A_c}{u_c^2} \sigma_c \dot{\theta} \sin\theta \qquad (8)$$

Where, $A_c$ is the frontal cross sectional area of the droplet, $A_s$ is the droplet surface area, D is the diffusion coefficient of the salt in water, $\Delta C$ is the difference in concentration between the bulk and at the interface (refer fig. 11), and $h(t)$ is the height of the droplet at the instant of analysis. The $u_c$ denotes the solutal Marangoni advection velocity and is scaled [49] as $u_c = \frac{\sigma_c \Delta C}{\mu}$. $\sigma_c$ is the rate of change of surface tension due to change in the salt concentration (obtained experimentally using the pendant drop method for a large number of salt concentrations and by fitting a best-fit correlation [55, 56]). Using the geometric parameters of the droplet, eqn. 8 is further expanded [49] as

$$\rho d_c^2 \dot{\theta} \sec^4(\theta/2) = 16 D \Delta C \cot(\theta/2) \left\{ Ma_s \sec^2(\theta/2) - 4 Ma_s \frac{Sc}{Ca_s} \sin\left(\frac{\theta}{2}\right) \cos\left(\frac{\theta}{2}\right) \right\} \quad (9)$$

Where, $Ma_s = \frac{\sigma_c \Delta C d_c}{2\mu D} \tan(\theta/2)$ is the solutal Marangoni number, $Ca_s = \frac{\mu d_c \dot{\theta}}{2\sigma} \sec^2(\theta/2)$ represents the solutal Capillary number, and $Sc = \frac{\mu}{\rho D}$ is the Schmidt number.

As per previous report by the authors [49], the solutal Marangoni advection plays a dominant in enhancing the evaporation dynamics over the thermal Marangoni advection. The relative strength of one over the other in case of elevated substrate temperatures needs to be mapped. It has been observed that that the increment in the values of thermal Marangoni number on heated substrate cases is large compared to the ambient case, but this does not prove dominance over the solutal Marangoni number. Figure 12 (a) and (b) illustrates the phase plot of the thermal Marangoni number with respect to the solutal counterpart on heated substrates. The stability criteria for the internal advection is represented by the iso-Le lines, as adapted from Joo [60]. All the points lying to the right side of the Le=0 line indicate stable circulation due to solutal Marangoni advection. It is observed that the points shift towards the right with increase in the substrate temperature, indicating enhanced circulation due to the solutal Marangoni advection in heated cases. However, comparison of the magnitudes shows that this increment is comparatively lesser than that observed for the thermal Marangoni number. The phase map thus shows that while the solutal Marangoni advection is dominant in the ambient evaporation case, the relative increase in strength of the thermal Marangoni advection is higher compared to the solutal counterpart in the case of heated substrates.

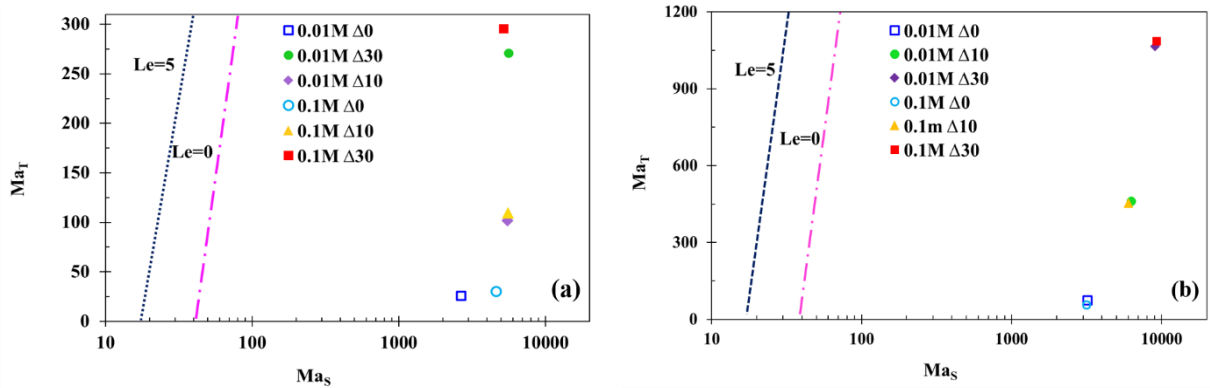

**Figure 12:** Phase plot of the thermal Marangoni with the solutal Marangoni number for saline droplets on different substrate temperatures for (a) hydrophilic substrate and (b) SHS. The stability curves denote iso-Le lines [60].

It is interesting to note that the thermal Marangoni advection analysis shows that the strength of the internal advection increases with increase in the substrate temperature. However, the velocimetry shows reduction in the internal circulation velocity. But it is also noteworthy that the same analysis presented was shown by the authors to predict the internal advection kinetics accurately [49]. Likewise, the solutal advection analysis reveals that the solutal advection strength is also augmented by increase in the substrate temperature, leading to increased circulation velocity. Thus, the analyses show that both thermal and solutal advection is augmented by the increased substrate temperatures. However, the velocimetry observations show that the internal circulation velocity is reduced in strength at higher substrate temperatures. It is proposed that the augmented thermal and solutal advection counter each other, which leads to reduction in the effective internal advection velocity. This theory is tested based on the present analyses. The velocity of thermal advection (from $u_M = \frac{\sigma_T \Delta T_m}{\mu}$) and that of solutal advection (from $u_C = \frac{\sigma_C \Delta C}{\mu}$) are independently deduced for the different experimental conditions. The effective average advection velocity within the droplet is theorized (from $|u_{eff}| = \frac{u_C - u_M}{2}$) as the competitive mean of the two independent velocities. The velocities predicted from the proposed theory are compared against the velocimetry observations and illustrated in figs. 7(c) and 8(c). It is noted that majority of the predicted velocities agree well with the experimental velocities, thereby confirming the proposed theory. Thereby, it is confirmed that the competitive thermal and solutal Marangoni advection leads to reduced internal advection, which leads to reduced evaporation rates of saline sessile droplets on heated surfaces.

## 4. Conclusions

To infer, the article sheds insight on the evaporation kinetics of sessile saline droplets on heated hydrophilic surfaces and SHS. The major points of highlight from this article are as follows:

- Experiments on sessile droplet evaporation of water and saline droplets at the ambient and different heated cases show that the evaporation kinetics possesses atypical and anomalous behaviour. In the ambient case, the saline droplets evaporate at faster rates compared to the water droplets.
- On the heated substrates, the water droplets evaporate faster than the ambient case. However, anomalously, the saline droplets evaporate slower compared to the water counterpart.
- Experiments show that the regimes of evaporation and their duration, along with the behaviour of the contact diameter and contact angles, are modulated by the substrate temperatures.
- PIV studies reveal that the internal advection in the water droplets is augmented by the substrate heating; however, the advection velocity is diminished in case of saline droplets on heated substrates. Thus the observations are counter-intuitive.

- Scaling analysis of the energy and species conservation reveals that the thermal and the solutal Marangoni numbers are augmented with substrate temperature. However, the effective improvement in the thermal Marangoni number is higher compared to that of the solutal Marangoni number. In essence, theory dictates that internal advection is augmented, whereas velocimetry reveals the opposite.
- It is theorized that the thermal and solutal Marangoni advection within such heated saline droplets are competitive in nature, and the opposing behaviour is effectively revealed experimentally. The velocity predictions from the theory are noted to be in good agreement with the velocimetry observations.

The article thus discusses anomalous and counter-intuitive evaporation behaviour of saline sessile droplets on heated surfaces, and puts forward the crux mechanism behind such characteristics. The findings may hold utilitarian implications in several micro and macroscale systems and applications employing hydrodynamics, and thermo-species transport phenomena in droplets, sprays and jets.